\documentclass[sigconf]{acmart}

\renewcommand\footnotetextcopyrightpermission[1]{} 
\usepackage{booktabs} 
\usepackage{amsmath}
\usepackage{diagbox}
\usepackage{amsmath}
\usepackage{amssymb}
\usepackage{subfig}
\usepackage{mathtools}
\usepackage{soul}
\DeclarePairedDelimiter\abs{\lvert}{\rvert}

\setcopyright{none}

\setcopyright{none}

\begin{document}
\title{A Broad Evaluation of the Tor English Content Ecosystem}

\author{Mahdieh Zabihimayvan}
\affiliation{%
  \institution{Department of Computer Science and Engineering\\ Kno.e.sis Research Center, Wright State University}
  \city{Dayton}
  \state{OH}
   \country{USA}
  \postcode{}
}
\email{zabihimayvan.2@wright.edu}

\author{Reza Sadeghi}
\affiliation{%
  \institution{Department of Computer Science and Engineering\\ Kno.e.sis Research Center, Wright State University}
  \city{Dayton}
  \state{OH}
   \country{USA}
  \postcode{}
}
\email{sadeghi.2@wright.edu}

\author{Derek Doran}
\affiliation{%
  \institution{Department of Computer Science and Engineering\\ Kno.e.sis Research Center, Wright State University}
  \city{Dayton}
  \state{OH}
   \country{USA}
  \postcode{}
}
\email{derek.doran@wright.edu}

\author{Mehdi Allahyari}
\affiliation{%
  \institution{Department of Computer Science \\ Georgia Southern University}
  \streetaddress{}
  \city{Statesboro}
  \state{GA}
   \country{USA}
  \postcode{}
}
\email{mallahyari@georgiasouthern.edu}



\begin{abstract}
Tor is among most well-known dark net in the world. It has noble uses, including as a platform for free speech and information dissemination under the guise of true anonymity, but may be culturally better known as a conduit for criminal activity and as a platform to market illicit goods and data. Past studies on the content of Tor support this notion, but were carried out by
 targeting popular domains likely to contain illicit content. A survey of past studies may thus
not yield a complete evaluation of the content and use of Tor. This work addresses this gap by presenting a broad evaluation of the content of the English Tor ecosystem. We perform a comprehensive crawl of the Tor dark web and, through topic and network analysis, characterize the `types' of information and services hosted across a broad swath of Tor domains and their hyperlink relational structure. We recover nine domain types defined by the information or service they host and, among other findings, unveil how some types of domains intentionally silo themselves from the rest of Tor. We also present measurements that (regrettably) suggest how marketplaces of illegal drugs and services do emerge as the dominant type of Tor domain. Our study is the product of crawling over 1 million pages from 20,000 Tor seed addresses, yielding a collection of over 150,000 Tor pages. 
We make a dataset of the intend to make the domain structure publicly available as a dataset at  \url{https://github.com/wsu-wacs/TorEnglishContent}.

\end{abstract}

%
%



\maketitle

\section{Introduction}
The deep web defines content on the World Wide Web that cannot or has not yet been
indexed by search engines. Services of great interest to parties that want to be anonymous 
online is a subset of the deep web called {\em dark nets}: networks running on the
Internet that require unique application layer protocols and authorization schemes to access. 
While many dark nets exist (i.e.
I2P~\cite{cyberSecurity}, Riffle~\cite{riffle}, and Freenet~\cite{freenet}), Tor~\cite{secondGeneration} has
emerged as the most popular~\cite{surfacing,onionEyes}. 
It is used 
as a tool for circumventing government censorship~\cite{censor1}, for releasing information to the public~\cite{censor2}, for sensitive communication between
parties~\cite{socialSciences, terrorism}, and as an (allegedly) private space to buy and sell goods and services~\cite{activism,weapons}.

It is an open question whether the fundamental and
often necessary protections that Tor provides its users is worth its cost: the same
features that protect the privacy of virtuous users also make Tor an effective means
to carry out illegal activities and to evade law enforcement. Various positions on this question
have been documented~\cite{marketplaceInvestigations,maliciousWebsites,organizedCrime},
but empirical evidence is limited to
studies that have crawled, extracted, and analyzed {\em specific subsets of Tor} based on 
the type of hosted information, such as 
drug trafficking~\cite{GeographicAnalysis}, homemade explosives~\cite{HybridFocused}, terrorist activities~\cite{Jihad}, or forums~\cite{socialSciences}. 

A holistic understanding of how Tor is utilized and its structure as an information ecosystem 
cannot be gleamed by surveying this body of past work. This is because previous studies 
focus on a particular subset of this dark web and take measurements that aim to answer unique 
collections of hypotheses. But such a holistic understanding of Tor's utilization and ecosystem 
is crucial to answer broader questions about this dark web, such as: \textit{How diverse is the information and the services provided on Tor? Is the argument that the use of
Tor to buy and sell illicit goods and services and to enable criminal activities valid?
Is the domain structure of Tor `siloed', in the sense that the Tor domain hyperlink network is highly modular conditioned on content? Does the hyperlink
network between domains exhibit scale-free properties?} Answers to such questions
can yield an understanding of the
kinds of services and information available on Tor, reveal the most popular and important
(from a structural perspective) services it provides, and enable a comparison of Tor against the surface web and other co-reference complex systems.

Towards this end, we present a quantitative characterization of the types of information available
across a large swath of English language Tor webpages. We conduct a massive
crawl of Tor starting from
20,000 different seed addresses and harvest {\bf only} the html page of each
visited
address\footnote{For privacy purposes no embedded resources of any kind, including images, scripts, videos, or other multimedia,
were downloaded in our crawl.}.
Our crawl encompasses over 1 million addresses, of which 150,473 are hosted on Tor and the remaining 1,085,960 returns to the visible web. We focus on 40,439 Tor pages
belonging to 3,347 English language domains and
augment LDA with a topic-labeling algorithm that uses DBpedia to assign
semantically meaningful labels to the content of crawled pages.
We further extract and study a logical network of English Tor domains connected by hyperlinks. 
We limit our study to the subset of Tor domains with information written in {\bf English} 
in an attempt to control for any variability in measurements and insights that could be caused by 
the confluence of information posted in unique languages, structure, and possibly unique subdomains
of Tor (for example, VPN services specifically targeting users in countries with government controlled censorship 
or marketplaces exclusively for users living in a particular country). 
We summarize our insights to the following research questions:
\begin{itemize}
	\item RQ1: {\em Is the information and services of Tor diverse?}
		\begin{itemize} \item We find that Tor services can be described by just nine types. Over 50\%
			of all domains discovered are either directories to other Tor domains, or serve as marketplaces to buy and sell goods and services. 
			Just 24\% of all Tor domains are used to publicly post, privately send, or to discover information anonymously. We do find, however,
			that different types of domains relate to each other in unique ways, including marketplaces that enable payment by forcing gameplay on 
			a gambling site, and that domains involving money transactions have a surprisingly weak reliance to Tor Bitcoin domains. 
			\end{itemize}
	\item RQ2: {\em What are the `core' services of Tor? Is there even a core?}
		\begin{itemize} \item An importance analysis from some centrality metrics finds the Dream market
			to be the most structurally important, ``core'' service Tor provides by a wide margin. 
			The Dream market is the largest marketplace for illicit goods and services on Tor. Directory sites to find and access 
			Tor domains have dominant betweenness centrality, making them important sources for Tor browsing.
		\end{itemize}
	\item RQ3: {\em How siloed are Tor information sources and services?}
		\begin{itemize} \item A connectivity analysis shows how 
			Tor services tend to isolate themselves which makes them difficult to discover by simple browsing and implies the need 
			for a comprehensive seed list of domains for data collection on Tor. 
			We also find patterns suggesting competitive and cooperative behavior between domains depending on their domain type, and that
			Tor is not particularly introspective: few news domains reference a large number of other domains across this dark web.
        \end{itemize}
	\item RQ4: {\em How can the structure of Tor domain hyperlinks be modeled? What
	are the implications of Tor's domain connectivity?}
		\begin{itemize} \item We find that the hyperlink network of Tor domains does not follow the scale-free structure observed in other 		
			sociotechnological systems or the surface web. Also, investigating within content communities denotes a power-tailed pattern 
			in the in-degree distributions of about half of all Tor domains.  \end{itemize}
\end{itemize}
To the best of our knowledge, this study reports on the largest measurements of Tor taken to date, and is the first to study the relationship
between Tor domains conditioned on the type of information they hold. We publish the hyperlink structure of 
this massive dataset to the public at: \url{https://github.com/wsu-wacs/TorEnglishContent}. Contrasting our findings for subsets 
of Tor written in other languages will be an exciting direction for future work. 

This paper is organized as follows: Section~\ref{sec:rw} discusses related work on
characterizing and evaluating Tor. Section~\ref{sec:dc} presents the procedure used for data collection and processing. Section~\ref{sec:tc} provides an evaluation on Tor content while Section~\ref{sec:dr} describes the logical network of the domains connected by hyperlinks and presents the analyses results. Finally, Section~\ref{sec:co} summarizes the main conclusions and discusses the future work.

\section{Related Work}
\label{sec:rw}
Previous research on Tor have focused on characterizing particular types
of hosted content, on traffic-level measurements, and on understanding the security, privacy, and topological
properties of Tor relays at the network layer. Towards understanding types of content on Tor, Dolliver {\em et al.} use geovisualizations and exploratory spatial data analyses to analyze distributions of drugs and substances advertised on the Agora Tor marketplace~\cite{GeographicAnalysis}. Chen {\em et al.} seek an understanding of terrorist activities by a method incorporating information collection, analysis, and
visualization techniques from 39 Jihad Tor sites~\cite{Jihad}.
M\"{o}rch {\em et al.} analyze 30 Tor domains to investigate the accessibility of information related to suicide~\cite{suicide}.
Dolliver {\em et al.} crawls {\em Silk Road 2} with the goal of comparing its nature in drug trafficking operations with that of the original site~\cite{silkRoad2}. Dolliver {\em et al.} also conduct an investigation on psychoactive
substances sold on {\em Agora} and the countries supporting this dark trade \cite{psychoactive}.
Other related works propose tools to support the collection of specific information, such as a
focused crawler by Iliou {\em et al.}~\cite{HybridFocused}, new crawling frameworks for Tor by
Zhang {\em et al.}~\cite{socialSciences}, and advanced crawling and indexing systems like LIGHTS
by Ghosh {\em et al.} \cite{AutomatedCategorization}.

Tor traffic monitoring is another related area of work. This monitoring is often done to
detect security risks and information leakage on Tor that can compromise the anonymity of its users
and the paths packets take.
Mohaisen {\em et al.} study the possibility of observing Tor requests at global DNS infrastructure that could
threaten the private location of servers hosting Tor sites, such as the name and onion address of
Tor domains~\cite{leakage}.
McCoy {\em et al.} study the clients using and routers that are a part of Tor by collecting
data from exit routers~\cite{shining}.
Biryukov {\em et al.} analyze the traffic of Tor hidden services to evaluate their
vulnerability against deanonymizing and take down attacks.
They demonstrate how to find the popularity of a hidden service, harvest their descriptors in a short time,
and find their guard relays.
They further propose a large-scale attack to disclose the IP address of a Tor hidden service~\cite{trawling}.
We consider such work, operating at the traffic level, as studying Tor payloads that pass through
networks, rather than in understanding the content of these payloads or of
the inter-connected structure of Tor domains.

\begin{figure}
\includegraphics[width=.48\textwidth]{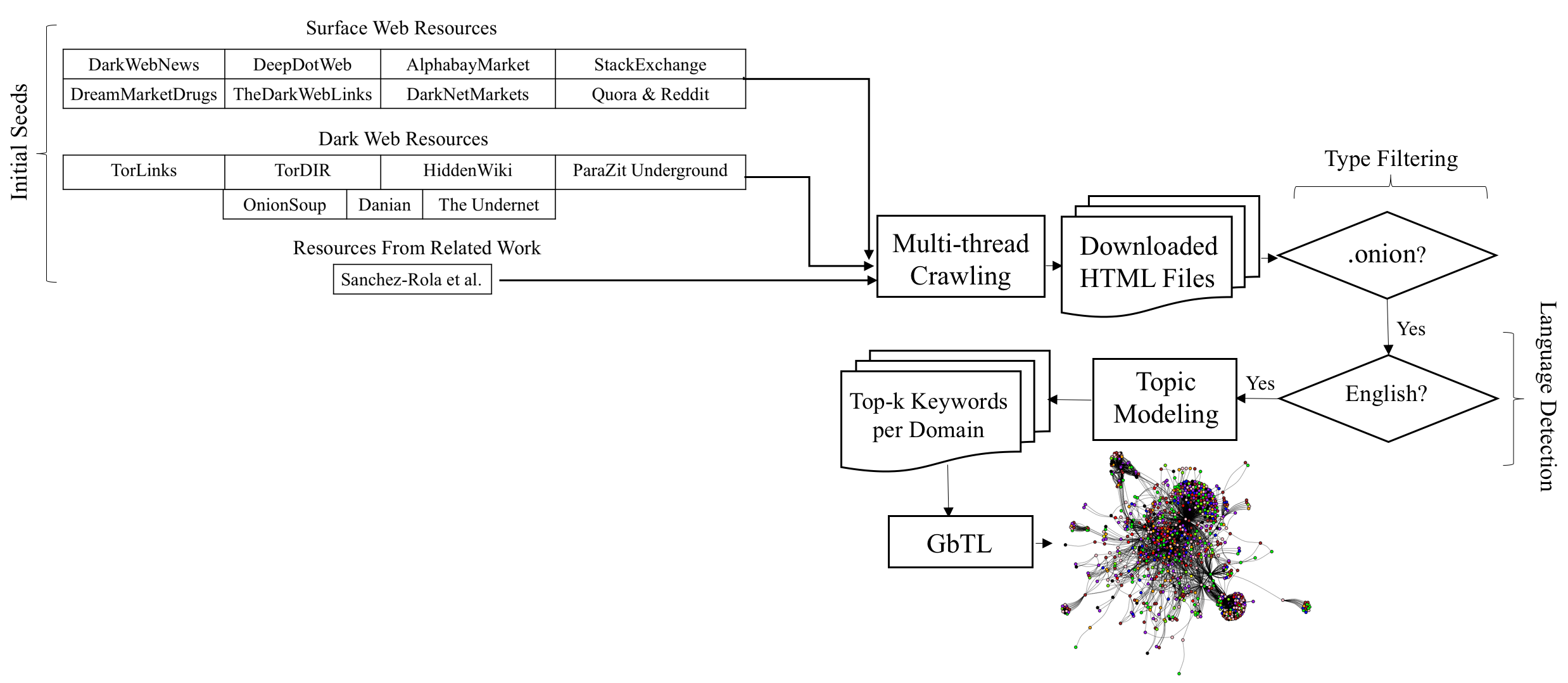}
\vspace{-15px}
\caption{Tor data collection process}
\label{fig:data_collection}
\vspace{-10px}
\end{figure}

The topological properties of Tor, at physical and logical levels, are only beginning to be
studied. Xu {\em et al.} quantitatively evaluated the structure of four terrorist and criminal related
networks, one of which is from Tor~\cite{topology}. They find such networks are efficient in communication and information flow, but are vulnerable to disruption by removing weak ties that connect large connected components.
Sanchez-Rola {\em et al.} conducted a broader structural analysis over 
7,257 Tor domains~\cite{onionEyes}.
Their experiments indicate that domains are logically organized in a sparse network, and finds
a surprising relation between Tor and the surface web:
there are more links from Tor domains to the surface web than to other Tor domains.
They also find evidence to suggest a surprising amount of user tracking performed on Tor. 
Part of this work extends their study by examining the structure between Tor domains 
{\em conditioned on the type of information they host}.

\section{Dataset collection and Processing}
\label{sec:dc}
We performed a comprehensive crawl of the Tor network to extract data for this
study. Figure~\ref{fig:data_collection} illustrates this data collection process.
We developed a multi-threaded crawler that collects the html of any Tor website
reachable by a depth first search (up to depth 4) starting from a seed list of
20,000 Tor addresses. This seed list was constructed by concatenating the
list used in a recent study~\cite{onionEyes} along those identified by the author's 
manual search of $Reddit$, $Quora$, and $Ahima$, and other major surface web directories in the days predating the crawl. 
Although a manual list of seeds runs the inevitable risk of a crawl that can miss portions of Tor,
the hidden nature of Tor websites make it unlikely for there to ever be a single authoritative directory
of Tor domains. We are confident that our seed list leads to a comprehensive crawl of Tor because:
~(i) The Reddit, Quora, Ahima, and surface web directories are well-known for providing up to date links to Tor domains
and are often used by Tor users to begin their own searches for information. This suggest that these entry points into 
Tor are at worst practically useful, and at best are ideal starting points to search and find Tor domains associated with 
the most common uses of the service;~(ii) The list adapted from~\cite{onionEyes} are noted to be sources commonly utilized to discover current Tor addresses. Furthermore, we assign our crawlers to cover all hyperlinks up to depth 4 from every seed page
to make our data collection as comprehensive as possible. Out seed list is published online.\footnote{\url{https://github.com/wsu-wacs/TorEnglishContent/blob/master/seeds}}

Because of the rapidly changing content and structure of Tor~\cite{onionEyes}, including temporary
downtime for some domains, we executed
two crawls 30 days apart from each other in June and July 2018. To try to control for some variability
in the up and down time of domains, the union of the Tor
sites captured during the two crawls were stored for subsequent analysis. 
It is worth noting that we only request html and follow hyperlinks, and do not download the full content of a web page. 
This prevents any access control polices, request rate limiters, and crawler blockers from interrupting our data collection.  
A total of 1,236,433 distinct pages were captured across both crawls.
The collected data was post-processed to identify English pages and to classify the crawled
pages as being from the surface or from Tor. Any webpage with suffix {\em .onion} was
classified as a Tor page, while the remainder are considered to be from the surface web.
A language identification method proposed by~\cite{textcat} was used to remove non-English onion pages based on their text content regardless of the value set in their HTML language tag.
40,439 English Tor pages remained after this filtering. We chose to only focus on English pages to facilitate our content analysis; an evaluation of non-English pages will be the topic of future work.

\subsection{Tor content discovery and labeling}
We subjected the corpus of English Tor pages through an unsupervised
content discovery and labeling procedure.
The process runs the content of every Tor page
(where content is defined as any string outside of a markdown tag)
through the Latent Dirichlet Allocation (LDA)~\cite{lda} and
Graph-based Topic Labeling (GbTL)~\cite{gbtl} algorithms
to  derive a collection of semantic labels representing broad topics of
content on Tor. Each Tor domain is then assigned a label by the dominant topic 
present across
the set of all webpages crawled in the domain.

\subsubsection{Topic Modeling}
Topic modeling \cite{topicModeling} is a method to uncover topics as hidden structures within a collection of documents. By defining a topic as a group of words occurring often together, topic modeling creates semantic links among words within the same context, and differentiates words by
their meaning. LDA ~\cite{lda} is a widely used unsupervised learning technique for this purpose. It
models a topic $t_j$  ($1 \le j \le T$) by a probability distribution
$p(w_i|t_j )$ over words taken from a corpus $D = \{d_1,d_2,\cdots,d_N\}$ of documents 
where words are drawn from a vocabulary $W = \{w_1, w_2,\cdots, w_M \}$.
The probability of observing word $w_i$ in document $d$ is defined as
$p(w_i|d)=\sum_{j=1}^{T}p(w_i|t_j)p(t_j|d)$. Gibbs sampling is used to estimate the word-topic
distribution $p(w_i|t_j)$ and the topic-document distribution $p(t_j|d)$ from data. 
\begin{figure}
\includegraphics[width=0.38\textwidth]{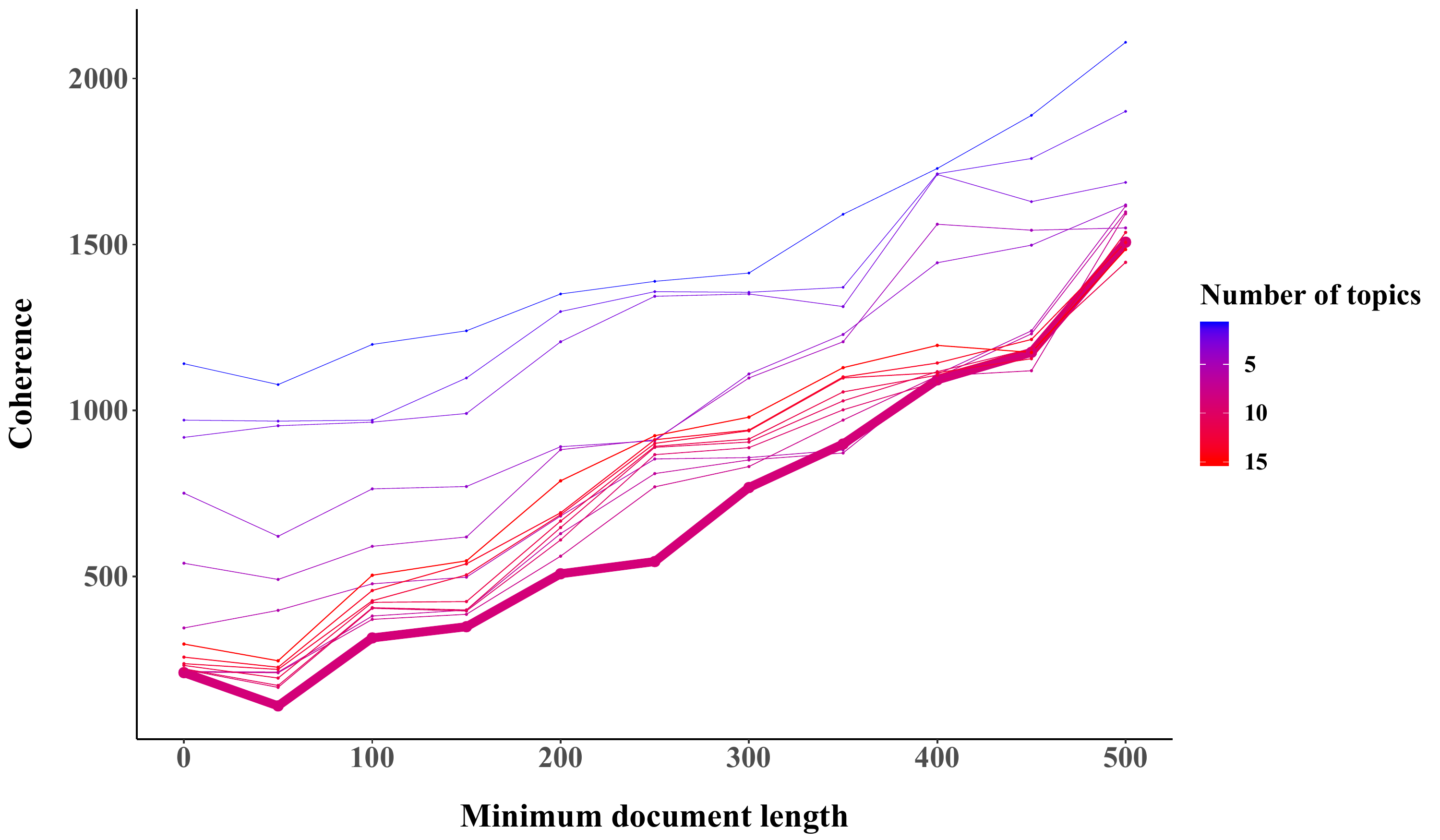}
\caption{LDA topic coherence score for different number of topic and minimum lengths of document; bold
trend representing scores using 9 topics}
\label{fig:KvsML}
\vspace{-15px}
\end{figure}

An important hyperparameter is the number of topics $T$ that should be modeled. We set $T$ by considering the {\em coherence}~\cite{McCallum} of a set of topics inferred for some $T$, choosing the $T$ with the best coherence $C$. 
$C$ is a function of the $n$ words of each $t_i$ having highest probability $P(w_j | t_i)$.
Let $W^{(t)}=\{w_1,\cdots,w_n\}$ be the set of top-$n$ most probable words from $P(w | t)$. Then $C$ is given by:
$$
C(t;W^{(t)})=\sum_{i=2}^{n}\sum_{j=1}^{i-1} log\frac{F_d(w_i^{(t)}, w_j^{(t)})+1}{F_d(w_j^{(t)})}
$$
where $w_i^{(t)}, w_j^{(t)} \in W^{(t)}$, $F_d(w)$ indicates the number of documents where $w$ emerges, and $F_d(w_i, w_j)$ gives the number of documents in which both words $w_i^{(t)}$ and $w_j^{(t)}$ exist.
Values closer to zero indicate higher coherence for the corresponding topic. 
$T$ is then chosen as the one that yields the smallest average $C$ over all topics: 
$$\arg\min_T \abs*{\sum_{t=1}^{T} \frac{C(t;W^{(t)})}{T}}$$ where the summation is ran over a model fitted to $T$ topics.

A final parameter is the minimum length of a document (e.g. Tor page) for it to be considered in the coherence
calculation. We determine this length empirically by inspection of Figure~\ref{fig:KvsML}, which gives topic coherence
scores for different values of $T$ and different minimum lengths of the documents where $n=10$ is considered for each topic. The trend for $T=9$ yields the closest value to zero for document minimum length of 50 words. 

\subsubsection{Graph-based Topic Labeling with DBpedia}
Word-topic distributions from a fitted topic model are indicative of semantically related words appearing in common contexts.
A human may subsequently assign a label to each topic by manually evaluating this distribution However, the manual approach yields a
subjective, possibly biased interpretation of the topics present in a corpus. Instead,
we incorporate the unsupervised knowledge graph-based labeling algorithm GbTL~\cite{gbtl}.
It utilizes the DBpedia~\cite{DBpedia} knowledge graph (KG) which codifies Wikipedia articles
and their relationships in the form of an ontology.
GbTL finds a concept $\zeta$ from DBpedia that would serve as a suitable label to represent $t$.

To determine $\zeta$, GbTL defines a suitability measure $\gamma$.
Before defining $\gamma$, we note that optimizing any measure over all DBpedia concepts is infeasible
due to the massive size of the DBpedia ontology.
Instead, GbTL considers a {\em candidate set} of possible labels for a topic $t$. The candidate 
set for topic $t$ are all vertices in the subgraph $G_t = (V_t, E_t)$
of DBpedia where $V_t$ is the set of concepts with labels identical to any word in $W^{(t)}$ along with their
directed 1st and 2nd degree neighborhood and $E_t$ is the relations among all concepts in $V_t$. 
The choice of 2nd degree neighbors is based on~\cite{gbtl} that found this setting to
produce a sufficiently large candidate set of labels without adding unrelated ones.
Since topic labels can be considered as assigning classes/categories to the topics, we restrict the
subgraph relations to those of type \textbf{rdfs:type}, \textbf{dcterms:subject}, \textbf{skos:broader}, \textbf{skos:broaderOf}, and \textbf{rdfs:sub-ClassOf}.
The suitability $\gamma$ of each concept $\zeta$ in $V_t$ is measured by its Focused Random Walk Betweenness Centrality measure~\cite{gbtl}. This centrality measures the average amount of time
it takes for a random walker to arrive at some node starting from any other node 
in a network. It is computed as follows:
\begin{enumerate}
\item Let $L=D-A$ be the Laplacian matrix of $G_t$ where $A$ is the adjacency matrix of $G_t$ and $D$ is its diagonal degree matrix.
\item Arbitrarily remove a row and its corresponding column from $L$ and then invert it.
Define $T$ as this inverse with a row and column vector of zeroes inserted at the same index
the row and column was removed from $L$. 
\item Define $\gamma(\zeta, t)$ as follows:

$$
 \gamma(\zeta, t)=
\begin{cases}
    \dfrac{\sum\limits_{v_x,v_y \in W^{(t)} , x < y}{I_i^{xy}} }{{n \choose 2}} ,& \text{if } \zeta \not\in W^{(t)}\\
    \dfrac{\sum\limits_{v_x,v_y \in W^{(t)} , x < y}{I_i^{xy}} }{\frac{(n-1)(n-2)}{2}},              & \text{otherwise}
\end{cases}
$$
\end{enumerate}
where $I_{i}^{xy}$ is given as:
$$
I_i^{(xy)}=\frac{1}{2}\sum_{v_j \in V_t} A_{ij} | T_{ij} - T_{iy}-T_{jx}+T_{jy}|
$$
where $i$ is the index of $A$ corresponding to $\zeta$ and $T_{ab}$ is the value at row $a$ and column
$b$ of $T$. 
Since DBpedia is a semantic graph, we assume that vertices playing an important structural role in the graph
should be representative of a concept that binds together the concepts in $W^{(t)}$. We use $\arg\max_\zeta \gamma(\zeta,t)$
as the label of topic $t$.

We built the topic models and derived their
subsequent labels using the 7,782 English Tor web pages having more than 50 words. 
These 7,782 pages come from 1,766 unique onion domains and represents the data we consider in the remainder of our study.\footnote{In the remainder of the paper, 
for sake of brevity we will sometimes refer to the set of English Tor domains as simply `Tor' or `Tor domains'.}
\begin{table*}
\caption{List of 10 most probable words per topic and their label}
\label{Keywords:label}
\begin{tabular}{ccl}
\toprule
Label& top-10 words\\
\midrule
Directory & Search, Database, Address, Tor, Browse, URL, Directories, Link, Site, Dir\\
Bitcoin & Btc, Bitcoin, Blockchain, Transaction, Wallet, Deposit, Coin, Buy, Anonymity, Hidden\\
News& Information, Newspaper, News, Tor, Events, Censorship, Web, Press, Tor, Comment\\
Email& PM, Privacy, Massage, Mail, GPG, Cypherpunk, AES, Webmail, IRC, Darkmail\\
Multimedia& Copyright, Book, Video, Music, Free, pdf, Library, TV, FLAC, Paper\\
Shopping& Supplier, Market, Hidden, Commerce, Product, buy, Order, Price, Marketplace, Money\\
Forum& NSFW, Invite, Friend, Group, Share, Private, BBS, IM, Chat, Forum\\
Gambling& Online, Gambling, game, Casino, Lottery, Roulette, Table, Value, Money, Play\\
Dream market& Marketplace, Online, Buy, Register, Dream, Support, Hidden, Account, User, Tor \\
\bottomrule
\end{tabular}
\end{table*}
\section{Content Evaluation}
\label{sec:tc}

\begin{figure}
\includegraphics[width=0.4\textwidth]{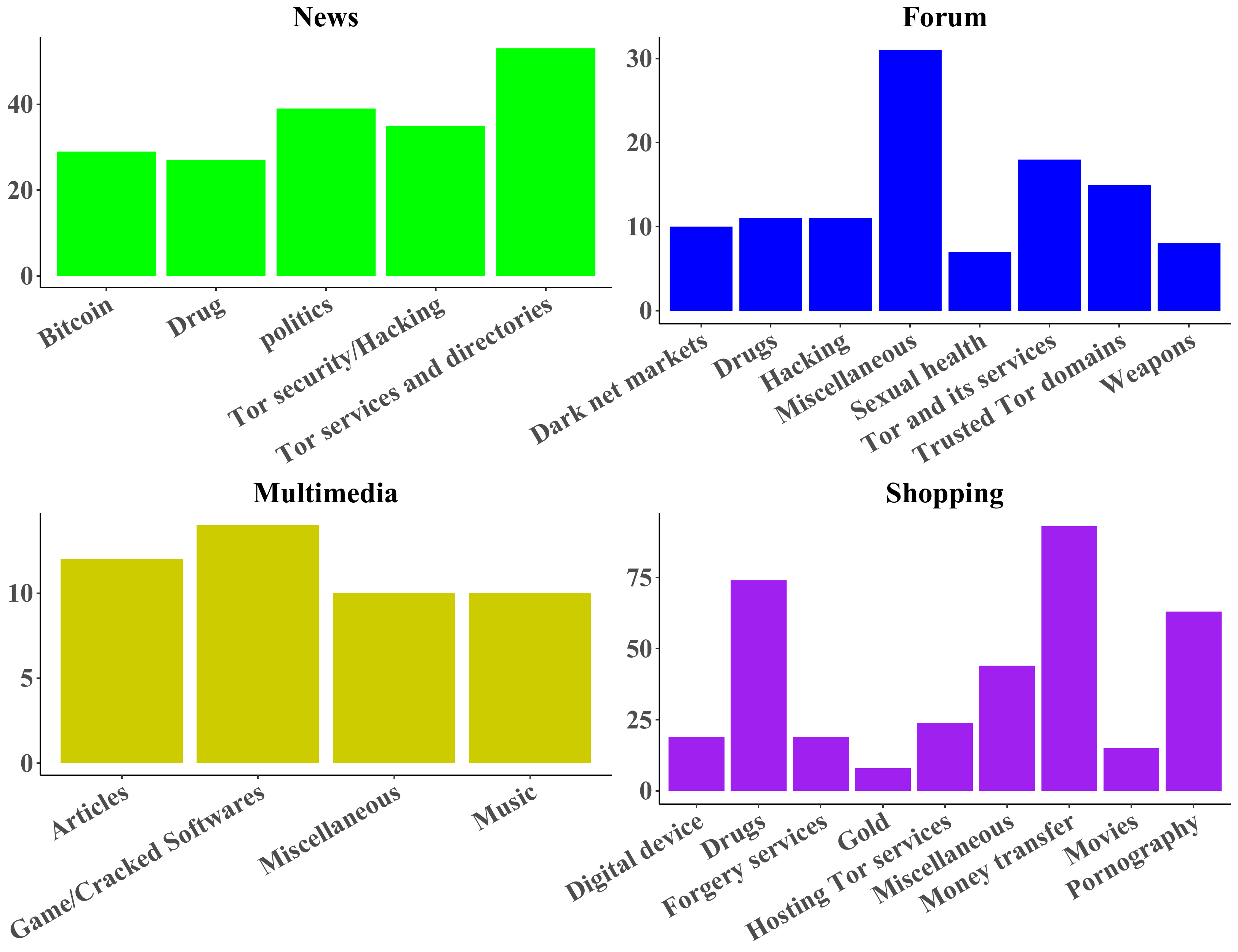}
\caption{Services provided by news, multimedia, forum, and shopping domains}
\label{fig:domainStats}
\vspace{-13px}
\end{figure}

We assigned each domain's label by the dominant topic in a concatenation of all 
of its constituent pages. 
Application of LDA to this set of documents yielded a set of 9 topics. GbTL
labeled these topics Forum, Shopping, Bitcoin, Dream market, Directory, Multimedia, News, Email service, and Gambling. 
Table~\ref{Keywords:label} lists the 10 most probable words per topic. We supplement this 
with a manual evaluation of the html code of pages in each domain to better understand the meaning of each
topic. 
For domains where we noticed a number of sub-types identified (e.g. a shopping
domain that specialized in particular types of goods or services), we maintained a 
count of sub-type frequencies and show them in Figure~\ref{fig:domainStats}. We elaborate on our
evaluation of each type of domain below: 

\noindent {\bf Shopping:} (359 domains [20.32\%]) 
Shopping domains allow visitors to purchase goods and services, including drugs,
medicine, as well as consultancy and investment services.
We found the dominant shopping service is to provide money transfer to and from credit cards
to buy Visa cards, gift cards, and Bitcoin. Drugs, pornography, hosting Tor services, and forgery services are other types of popular shopping domains. The rest includes pages selling a variety of goods like phone, laptop, movies, and even gold.
		   
\noindent {\bf Bitcoin:} (83 domains [4.7\%]) 
Bitcoin domains provide services for Bitcoin transactions and fund transfers to wallets. 
Tor Bitcoin services differ in terms of the registration process and personal information needed, registration and transaction fees, 
the security platforms they use,
and fees compared to surface web counterparts. 

\noindent {\bf Dream market:} (235 domains [13.3\%]) 
One shopping domain so large that it merited its own topic category is the Dream market. 
Its pages suggest a wide range of content available for
sale, most of which are illicit. This includes drugs, stolen data, and counterfeit consumer goods.
Many Dream market pages collected includes login and registration forms (hence words like ``Register'', ``Account'',
and ``User'' emerge in the list of top-10 terms in Table~\ref{Keywords:label}) limiting further access for our crawlers.

\noindent {\bf Directory:} (445 domains [25.19\%])
Addresses in Tor are made up of meaningless combinations of digits and characters that sometimes change and are hard to memorize.
Moreover, most domains may disappear after a short amount of time or move to new addresses \cite{onionEyes}. It is
thus no surprise to find directory domains emerge in our study. Directories are a convenient way of finding Tor domains without knowing their addresses. Domains labeled as a directory include unnamed pages with lists of \textit{.onion} domains along with the better known \textit{TorDir} and \textit{The Hidden Wiki} services, as well as search engines like \textit{DeepSearch} and \textit{Ahmia}.

\noindent {\bf Multimedia:} (46 domains [2.6\%]) 
Multimedia domains are sites to download and 
purchase multimedia products like e-books, movies, musics, games, and academic and press articles even if they are copyright protected. Among all multimedia domains, 
we measured 28\% of them to exclusively offer articles and e-books while 22\% offer free download of music or audio files, and to even obtain login information for stolen TV accounts. The remaining provides resources like hacked video game accounts,
cracked software, and a mixture of the above.

\noindent {\bf Forum:} (111 domains [6.3\%])
Forum domains host bulletin board and social network services for Tor users to discuss ideas and thoughts.
Among all forums, 72\% have a range of topic discussions including information on Tor and its services, hacking, sexual health,
dark net markets, weapons, drugs, and trusted Tor domains. Some forums require payment via Bitcoin to register.

\noindent {\bf News:} (183 domains [10.36\%])
Whereas forums facilitate interaction among Tor users, news domains host pages 
akin to personal weblogs where an author writes an essay and visitors can post follow-up comments.
Links to Tor e-mail services are included to contact post authors. Information on current Tor 
services and directories is presented by most news domains, along with politics, Bitcoin, drug, and Tor security related posts.

\noindent {\bf Email service:} (124 domains [7.02\%])
Email domains offer communication services like email, chat room, and Tor VPNs. Email services vary in the encryption protocol they use, the advertisements they serve for other services, and policies for keeping user log files. Our investigation finds that many email domains use secure protocols like SSL, AES, and PGP to secure user accounts and messages. IRC-based chat rooms are also common. Some email services charge recurring subscription fees.

\noindent {\bf Gambling:} (180 domains [10.19\%])
Gambling domains offer services to bet money on games, to purchase gambling advice and consulting, and to read
gambling-related news. Gambling domains have a number of links to payment processing options including
Ethereum, Monero, DASH, Vertcoin, Visa, and MasterCard.
\begin{figure}
\includegraphics[width=0.3\textwidth]{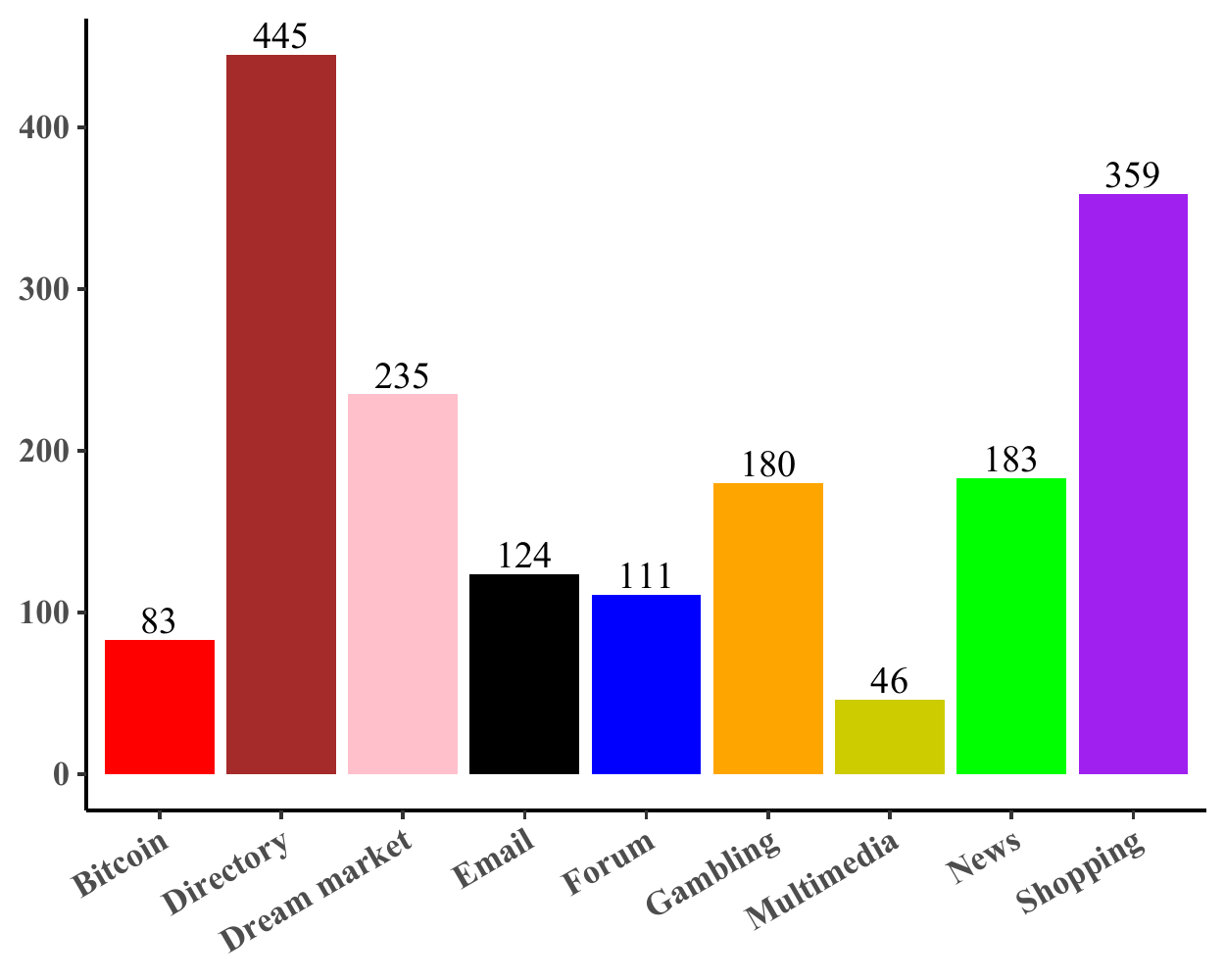}
\caption{Topic distribution of Tor domains}
\label{fig:tdist}
\vspace{-10px}
\end{figure}

Figure~\ref{fig:tdist} gives the distribution of topics assigned to each Tor domain. It illustrates
how directory and shopping domains, and the Dream market dominate English domains on Tor by
accounting for 58.83\% of all domain types. This suggests that Tor's main utility for users may be to browse information and shop on marketplaces that
require secrecy. In contrast, domains related to the free exchange of ideas and information
(a powerful and positive use-case of Tor, particular to users in countries facing Internet censorship~\cite{censor2}),
represented by Forum, Email, and News sites, account for just 23.66\% of all domains. It should be noted, however that
services used in countries that most benefit from the freedom of expression of Tor may not be well-represented
in our sample. Gambling domains represent a surprisingly large (10.19\%) percentage of domains, suggesting that people may now be
turning to Tor to play online gambling games that are otherwise illegal to host in many countries around the world.
Finally, and perhaps surprisingly, Bitcoin and multimedia domains respectively constitute the smallest proportion of
English Tor domains. Our initial expectation was that Bitcoin domains would be popular given the prevalence of the
Dream market and shopping domains where purchases are made via cryptocurrency. We postulate that sites having a Bitcoin
domain may host wallets, search a blockchain, or be markets that covert currency to BTC. Such sites hence may only
need to be visited infrequently. Moreover, the mainstream popularity of Bitcoin has led to 
reputable surface web domains offering similar Bitcoin services. 

\section{Domain Relationships}
\label{sec:dr}
We next examine the {\em structure of relationships} between different sources of information on Tor.
Such structural analyses realize the inter and intra-connectivity of domains on Tor conditioned
by the type of information or content they host. The structural analysis also seeks to identify the 
topological properties of the Tor domain network to evaluate similarities and differences
between the structure and formation process of Tor domains compared to the surface web and other 
sociotechnological systems. We build a graph where vertices are domains and a directed relation means
a page in a domain has a hyperlink to a page in a different domain. 

\begin{table}
\caption{Summary statistics of the domain network}
\begin{tabular}{lc}
\toprule
Statistic & Value \\
\midrule
Domain count ($|V|$) & 1,766\\
Hyperlink count ($|E|$) & 5,523\\
Mean degree ($\bar{k}$) & 12\\
Max degree ($\max{k}$) & 389\\
Density ($\rho$) & 0.0064\\
W.C.C. Count ($|C^w|$) & 25\\
S.C.C. Count ($|C^s|$) & 756 \\
Max W.C.C. Size ($\max{C^w_i}$) & 955 (54\%)\\
Max S.C.C. Size ($\max{C^s_i}$) & 13 (0.73\%)\\
\bottomrule
\end{tabular}
\vspace{-20px}
\label{tab:stat}
\end{table}

\begin{figure}
\includegraphics[width=0.35\textwidth]{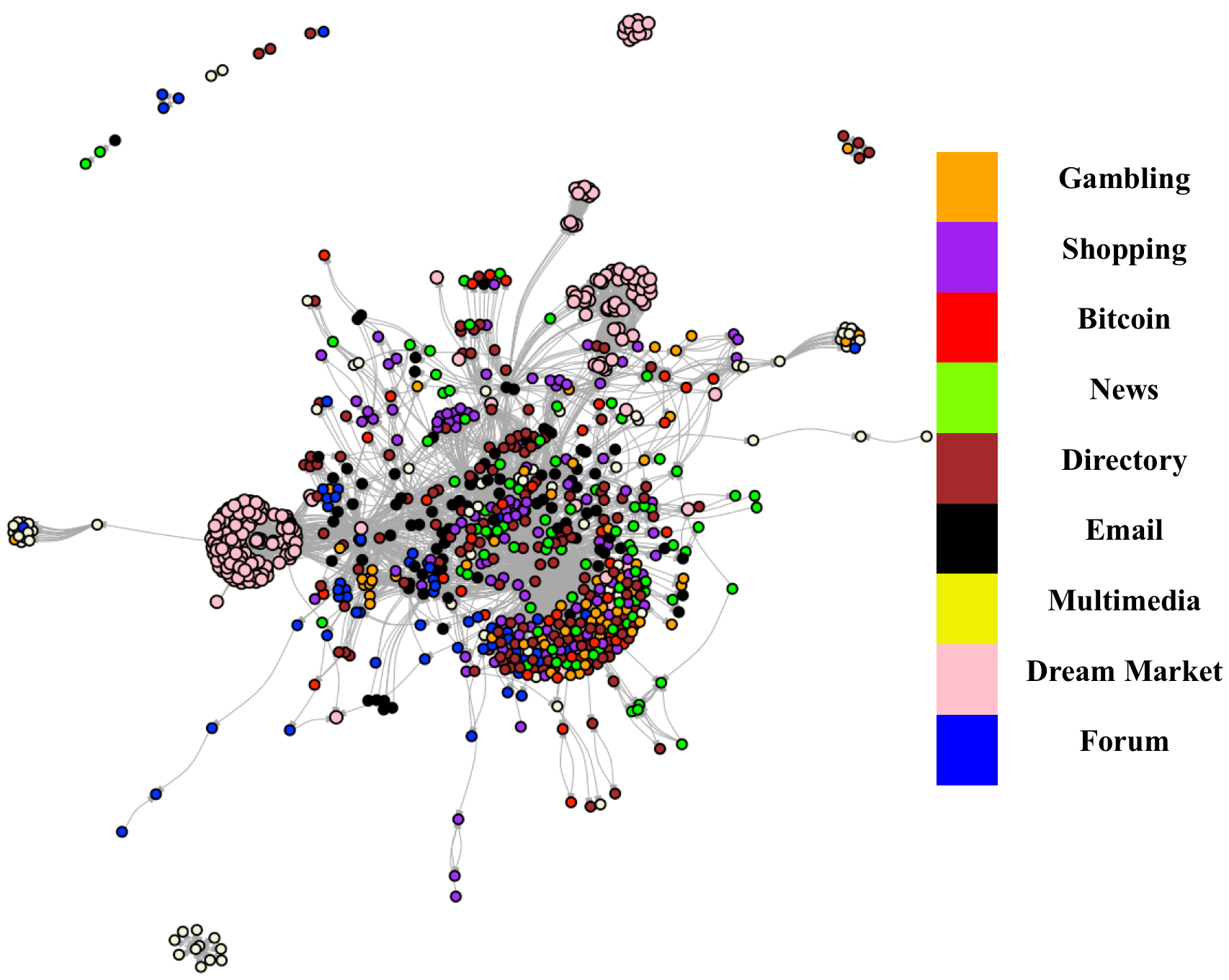}
\caption{Network of domains with degree > 0}
\label{fig:network}
\vspace{-15px}
\end{figure}
We measure simple statistics on the 
connectivity and connected components of the graph in Table~\ref{tab:stat} 
to make sense of the structure visualized in Figure~\ref{fig:network} (note that all  
domains with degree $0$ are excluded from the figure).
The network is sparse, with only 5,523 undirected edges among 1,766 domains and a network density $\rho=0.006$. 
The network also has $|C^w| = 25$ weakly connected components (W.C.C.) with the largest one having 955 domains. That only 54\% of all
vertices fall in the largest W.C.C. is somewhat surprising as many sociotechnological systems including the
surface Web hyperlink graph exhibit a single massive connected component that the vast majority of all vertices
participate in~\cite{connectedWeb}. 
This suggests that the underlying process for linking between websites is fundamentally different 
than in most sociotechnological systems: rather than encouraging connections to make information dissemination 
easier, the modus operandi of Tor domain owners may be to discourage linking to other domains, so that information on this 
``hidden'' web becomes difficult to arbitrarily discover and disseminate.
 This hypothesis is further supported
by examining the set of strongly connected components $C^s$. 
We measure $|C^s| = 756$ and $\max{C^s_i} = 13$, suggesting that an extraordinarily small number of 
domains collectively co-link with each other. 
It may be the case that the relatively larger $|C^w|$ and $\max{C^w_i}$ are simply caused by directory 
websites that offer links to a variety of other domains. These interesting deviations from 
other types of Web graphs collectively suggest that information on Tor may be intentionally isolated 
and difficult to discover by simple browsing.  
The small maximum connected component size speaks 
to the need of a 
comprehensive seed list of domains to for any comprehensive crawl of Tor. 

\subsection{Connectivity analysis}
\label{sec:conn}
We next examine hyperlink relationships across domains, within domains, and 
the tendency of domains to link to like domains.

\begin{figure*}
	\centering
\includegraphics[width=0.65\textwidth]{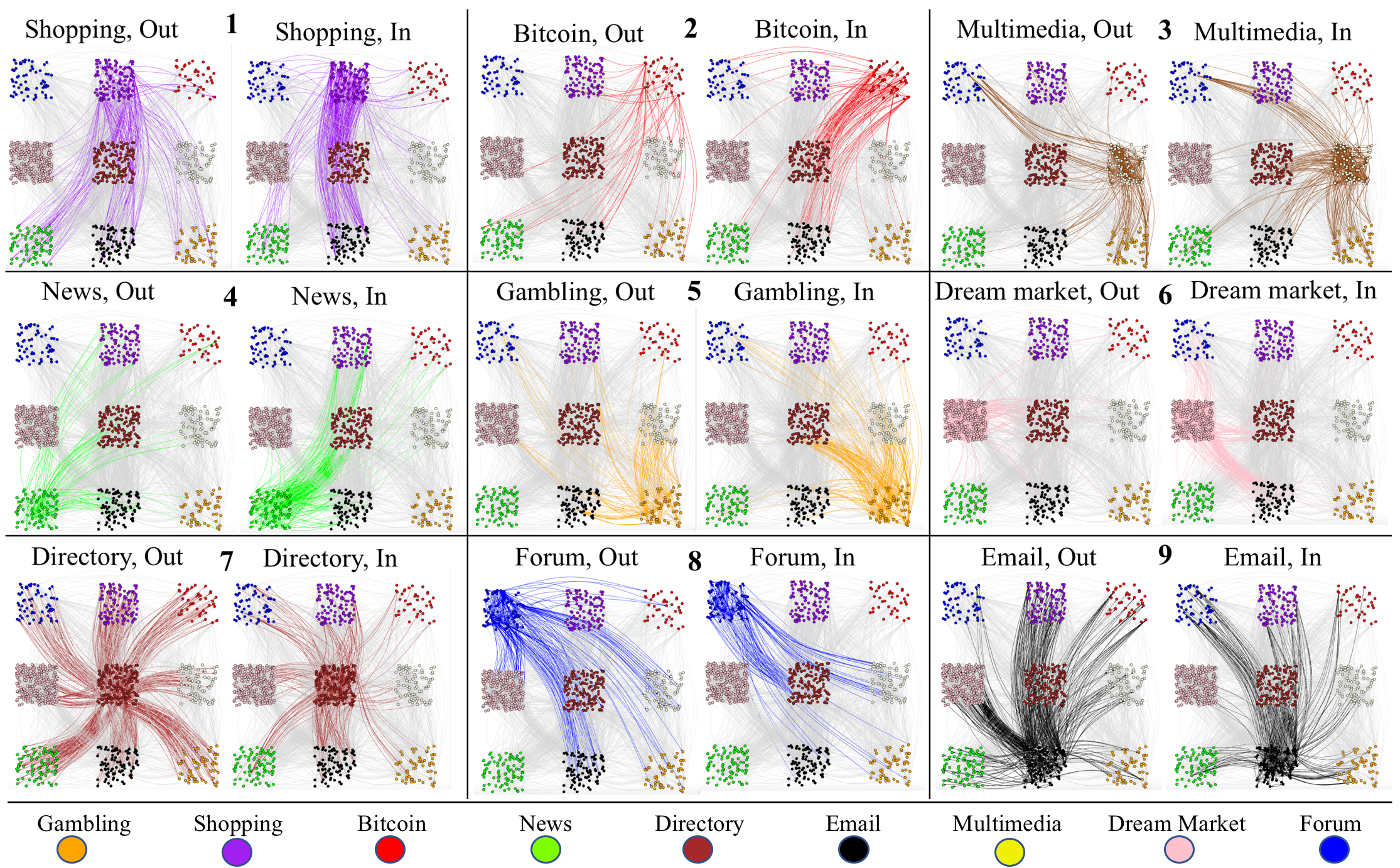}
\caption{The Tor domain network. Panels 1 through 9 each highlight the incoming and outgoing edges of a particular domain.
The figure is best viewed digitally and in color.}
\label{fig:carpet}
\end{figure*}

\begin{figure}
\includegraphics[width=0.48\textwidth]{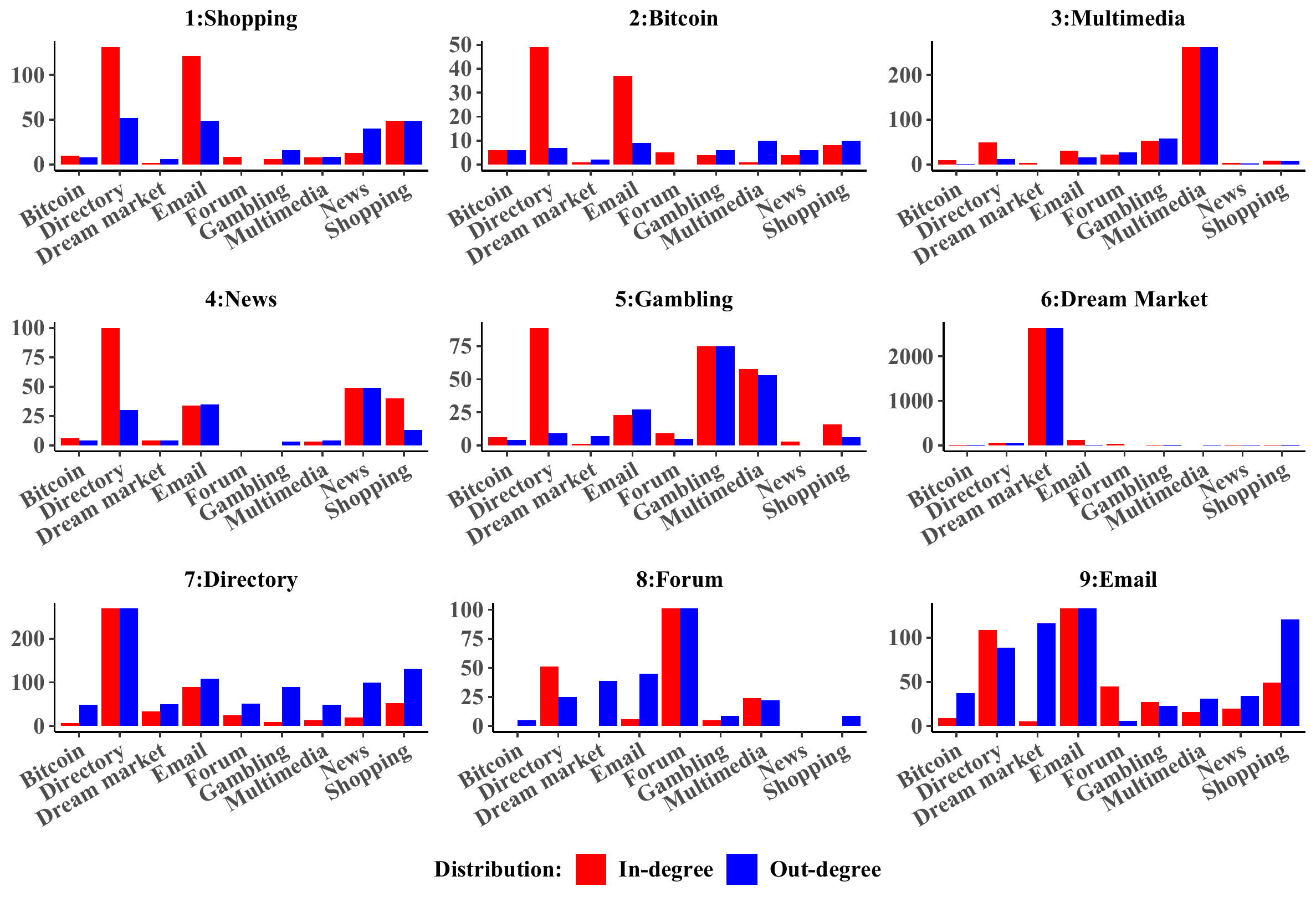}
\caption{In/out degree distribution of each community}
\label{fig:deg}
\vspace{-10px}
\end{figure}

\subsubsection{Inter-connectivity}
\label{sec:interConn}
To investigate the relationship between specific domains, 
we redraw the network with a carpet layout where vertices are spatially grouped 
in a grid by their domain type in Figure~\ref{fig:carpet}. We also list the sum of in- and out-degree from each community to every 
other community in Figure~\ref{fig:deg}. 
There are some notable domain relationships observed in the two figures.
For example, we find a small number of outgoing edges from shopping to gambling domains 
(Panel 1 of Figures~\ref{fig:carpet} and~\ref{fig:deg}). 
Our manual investigation of these relations find that some shopping domains actually provide customers with a 
method of payment by gambling, where a customer and seller play an online game to determine an amount of payment. 
We further note that shopping websites are isolated from the Dream market, perhaps to maintain some distinction between their offerings and the largest marketplace on Tor~\cite{dreamM}.
Another interesting observation is that there are a few number of edges from shopping to Bitcoin domains. Our manual investigation indicates that in addition to some shopping domains providing in-person cash payment (usually local drug vendors), marketplaces that use
cryptocurrency for payment tend to link to providers on the surface web, sometimes with instructions for the user to purchase
cryptocurrency from a trading house or market, thus explaining the small number of out-going edges from marketplaces to Bitcoin domains. 
Such links to the surface web, however have been noted as a major Achilles heel to privacy on
Tor as a cause of Tor information leakage~\cite{onionEyes}.
Links incoming to Bitcoin domains
(Panel 2 of Figures~\ref{fig:carpet} and~\ref{fig:deg}) are dominated by
directory ($\approx 53\%$) and email ($\approx 32\%$) domains.
This finding defeats our intuition that marketplaces, the Dream market, 
and gambling sites would have been services
most reliant on Bitcoin domains. We also investigated the email domains having edges to Bitcoins and found that many email services on Tor charge customers for anonymous messaging services or suggest a donation be made via Tor Bitcoin services. 
The sparse relation between multimedia and Bitcoin
domains are due to the fact that some multimedia domains charge users for the services they provide, and of those, many utilize surface web cryptocurrency providers.

Other insights can be gleamed from the carpet and inter-domain degree
distributions. 
For example, the outgoing connections from
Dream market websites (Panel 6 of Figures~\ref{fig:carpet} and~\ref{fig:deg}) show that 
this marketplace has very few incoming or outgoing edges to
other domains. Incoming links tend to originate from directory, email, and forum communities, which could be a byproduct of forum threads and email links to this secret marketplace. The Dream market thus appears to be especially siloed on the dark web, despite the fact that it represents $13.3\%$ of all domains. 
The outgoing edges from directory domains (Panel 7 of Figures~\ref{fig:carpet} and~\ref{fig:deg}) indicate that a large
collection of directories on Tor may be sufficient to include sites across all 
major Tor domains. Email domains exhibit a similar phenomena (Panel 9 of Figures~\ref{fig:carpet} and~\ref{fig:deg}) where they
tend to connect to all other types of domains. This suggests that Tor e-mail services
are not necessarily exclusive to only marketplace or forum users, but may be
a useful service for most Tor visitors. Finally, we see that news and forums have no hyperlinks to each other (Panel 4 of Figures~\ref{fig:carpet} and~\ref{fig:deg} and Panel 8 of Figures~\ref{fig:carpet} and~\ref{fig:deg}) which implies that although both services are information providers on Tor, they work independently of each other.

\subsubsection{Intra-connectivity}
\label{sec:intraConn}
\begin{figure}
\includegraphics[width=0.4\textwidth]{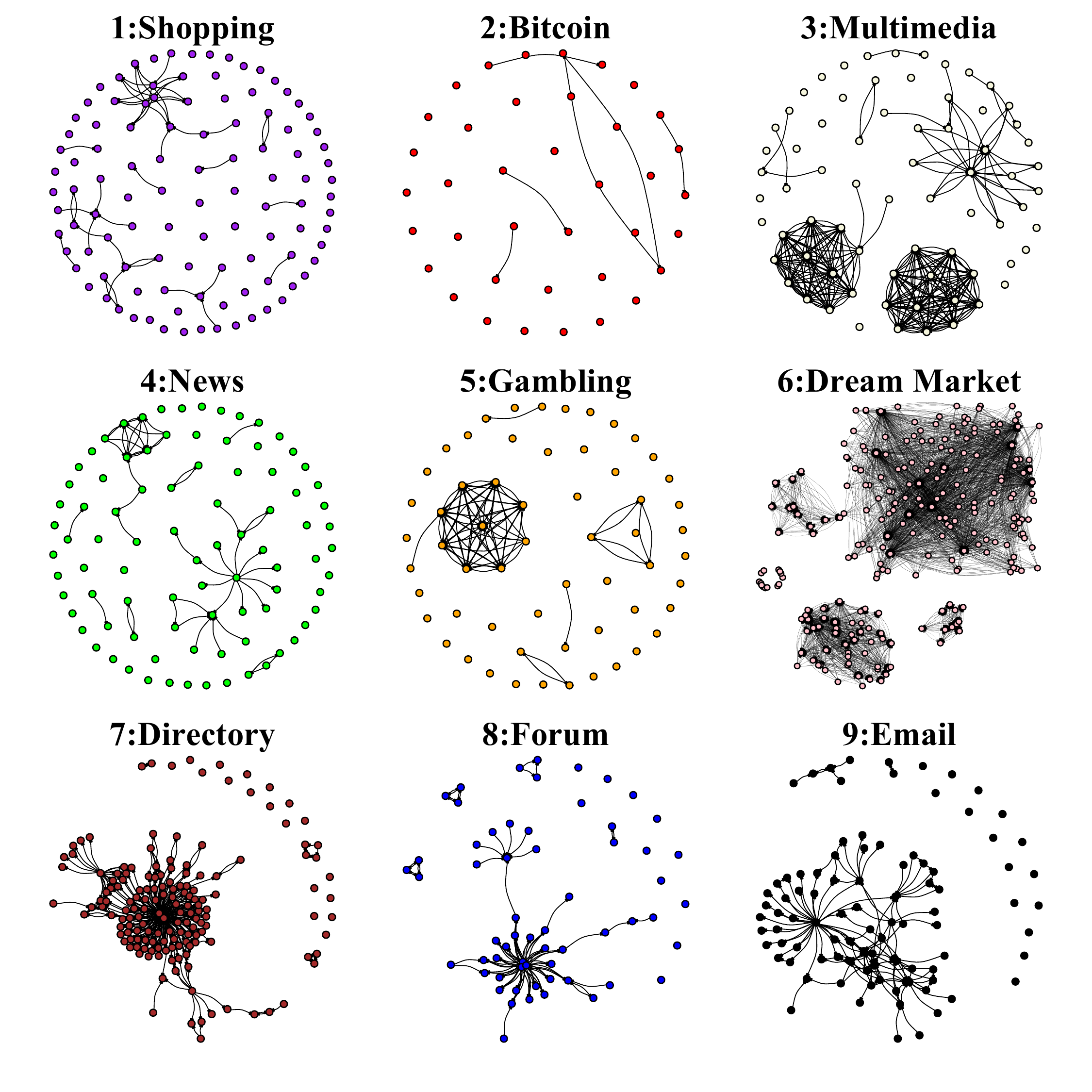}
\caption{Intra-relations within domains}
\label{fig:communityRels}
\vspace{-15px}
\end{figure}

Next, we investigate the connectivity within each domain to evaluate how tightly knit and accessible particular types of Tor
domains are among each other. This can reveal how often domains encourage their visitors to visit other domains having similar
types of content. Figure~\ref{fig:communityRels} separately visualizes all intra-domain connections for each domain type. 
Perhaps unsurprisingly, Dream market domains are tightly connected compared to others, splitting into only four connected components 
and the smallest number of isolated domains. Shopping domains are almost entirely disconnected from each other, indicating that  
marketplaces may intentionally disassociate themselves with others. Gambling, multimedia, and Bitcoin domains are similarly disconnected,
likely reflecting a competition for users within domains that tend to charge service fees.
On the other hand, email, forum, and directory domains all exhibit a single large connected component. 
This implies that in email and directory communities, domains have more support from each other and refer their visitors to other 
similar service providers. News domains have a larger percentage of isolated domains which implies that most of the news websites work 
independently from each other.

\begin{table}[h]
\caption{$R_\kappa$ for Tor domain networks}
\centering
$\begin{array}{ *{4}{l} }
\toprule
\textup{Community} & R_b & R_c & R_d \\
\midrule
\textup{Shopping}& .07 & .04 & .03\\[-1pt]
\textup{Bitcoin} & .09&.08& .08\\[-1pt]
\textup{Multimedia}& .22&.12&.11\\[-1pt]
\textup{News}& .10& .05&.04\\[-1pt]
\textup{Gambling}& .11&.06&.06\\[-1pt]
\textup{Dream Market} & .10&.35&.03\\[-1pt]
\textup{Directory}& .17&.11&.02\\[-1pt]
\textup{Forum}& .24&.14&.09\\[-1pt]
\textup{Email}& .52&.18&.05\\[-1pt]
\bottomrule
\end{array}$
\label{tab:R-coef}
\end{table}

We also quantify the intra-connectivity of topic domains by a Robustness coefficient $R_\kappa$ proposed by~\cite{Rcoef}. 
This coefficient reflects the degree to which a network shatters into multiple connected components as vertices and their incident
edges are removed. 
To compute $R_\kappa$, an ordering $\mathcal{O}_\kappa$ is induced on the vertices of 
the network by a centrality measure $\kappa$ such that $\mathcal{O}_\kappa(k)$ gives the vertex with 
the $k^{th}$ highest $\kappa$ centrality. Letting $C_i^{(\kappa)}$ be the size of the 
largest connected component of the network after removing 
$\{\mathcal{O}_\kappa(1), \mathcal{O}_\kappa(2), ..., \mathcal{O}_\kappa(i)\}$ and their incident edges,
$R_\kappa$ is given by: 
$$
R_\kappa = \frac{S_1}{S_2}=\frac{\sum_{i=0}^{|V|} i \cdot C_i^{(\kappa)}}{\sum_{i=0}^{|V|} i|V|-\sum_{i=0}^{|V|} i^2}=\frac{6\sum_{i=0}^{|V|} i \cdot C_i^{(\kappa)}}{|V|(|V|+1)(|V|-1)}
\label{equ:R}
$$
$R_\kappa$ ranges in $[0,1]$ and smaller values suggest that the network shatters faster
as vertices having high $\kappa$ betweenness are removed. 
The idea behind $R_\kappa$'s formulation is to quantify the change in the largest network 
component size $C_i^{(\kappa)}$ as nodes are removed from the network. 
The network is `maximally robust' if a plot of $C_i^{(\kappa)}$ vs the number of nodes removed shows
a simple linear decreasing trend. Then the ratio of area under such a plot for a given network, $S_1$, 
over the area under the plot for an ideal network, $S_2$, defines the robustness coefficient for that network.
Technical details about the measure are discussed in~\cite{Rcoef}.
Table~\ref{tab:R-coef} lists $R_\kappa$ for each 
intra-domain network under Betweenness $b$, closeness $c$, and degree $d$ centrality. 
$d$ is defined as the undirected degree of a node, $b$ is defined as the number
of shortest paths in the network that a node participates in, and $c$ is defined
as the inverse of the average path length from the node to all others in its connected
component~\cite{centrality}.

Studying Table~\ref{tab:R-coef} shows a correlation between $R_\kappa$ and the competitive or cooperative intra-domain behaviors noted above. 
There are some networks whose $R_d$ is close to $R_c$, namely those communities that are sparse networks (shopping, Bitcoin, news, gambling, multimedia), where competition for users and attention is natural. For domains where $R_c$ and $R_d$ are differentiated (directory, email, forum, Dream market), 
their structure has fewer connected components. This is best illustrated by Dream market 
where the difference of these coefficients has its maximum value, while exhibiting the fewest connected components. Also, we see that 
for networks with more supportive interlinking among domains (directory, email, and forum) the difference between $R_c$ and 
$R_d$ are larger.

In comparing $R_b$ between domains, we observe two groups with similarly small (shopping, Bitcoin, news, Dream market, gambling) and large (multimedia, directory, forum, email) values. This suggests that the intra-connectivity of domains in 
the first group is dependent on a small percentage of domains, or has a high number of isolated domains which have zero Betweenness centrality. On the other hand, high $R_b$ in the second category implies that their intra-connectivity is more robust to domain failures. 

It is worth mentioning that for Dream market, $R_b$ and $R_c$ are significantly different due to the separated connected components also seen in Figure~\ref{fig:network}. Since a high percentage of domains in this community are located in a single 
connected component, their closeness centralities will be greater than zero. On the other hand, the separated connected components can cause low values for Betweenness centrality since this metric is based on paths between pairs of domains. 
This finding also indicates that based on the information we have from home pages of Dream markets, there exist separated Dream market communities that link to similar types of goods and services.

\subsubsection{Modularity}
\label{sec:modul}
Finally, we study the modularity of the network as a means to understand the relationship between the 
inter- and intra-domain connectivity conditioned on the content type of the domain. 
A domain will have high modularity if it tends to link to pages of the same content type, and low modularity
if it tends to link to pages of a different type. The modularity $M$ of a network is given as~\cite{centrality}: 
$$
M=\frac{1}{2|E|}\sum_{i,j} \left(A_{ij}-\frac{d_id_j}{2|E|}\right)\Delta_{type_i = type_j}
$$
where $A$ is a binary network adjacency matrix with $A_{ij} = 1$ if $v_i$ and $v_j$ are adjacent and 
$d_i$ and $d_j$ are the undirected degree of nodes $i$ and $j$, and $\Delta_{E}$ is an 
indicator that returns $1$ if statement $E$ is true and $0$ otherwise.
Table~\ref{tab:modularity} shows that Dream market domains exhibit highest modularity by a wide margin, 
reinforcing the idea that Dream market domains are largely siloed from all other domains in the Tor ecosystem.  
Directories have lower but non-negligible modularity, leaving the impression that directory domains weakly cooperate
by linking to each other. The remaining domains have substantially lower modularity; thus the majority of 
Tor domains strongly prefer to link to other types. This suggests that Tor domains
prefer to remain isolated within their community of like domains, electing not to link to domains
that offer the same type of information or services. 

\begin{table}
\vspace{-5px}
\caption{Modularity score of each topic community}
\begin{tabular}{lc}
\toprule
Community & $M$\\
\midrule
Dream market & 0.452\\[-1pt]
Directory & 0.068\\[-1pt]
Forum & 0.034\\[-1pt]
Email & 0.032\\[-1pt]
Gambling & 0.024\\[-1pt]
News& 0.019\\[-1pt]
Multimedia & 0.017\\[-1pt]
Shopping & 0.012 \\[-1pt]
Bitcoin & 0.002\\[-1pt]
\bottomrule
\end{tabular}
\label{tab:modularity}
\vspace{-10px}
\end{table}

\begin{figure}[ht]
\vspace{-15px}
\subfloat[Betweenness]{\includegraphics[width=0.15\textwidth]{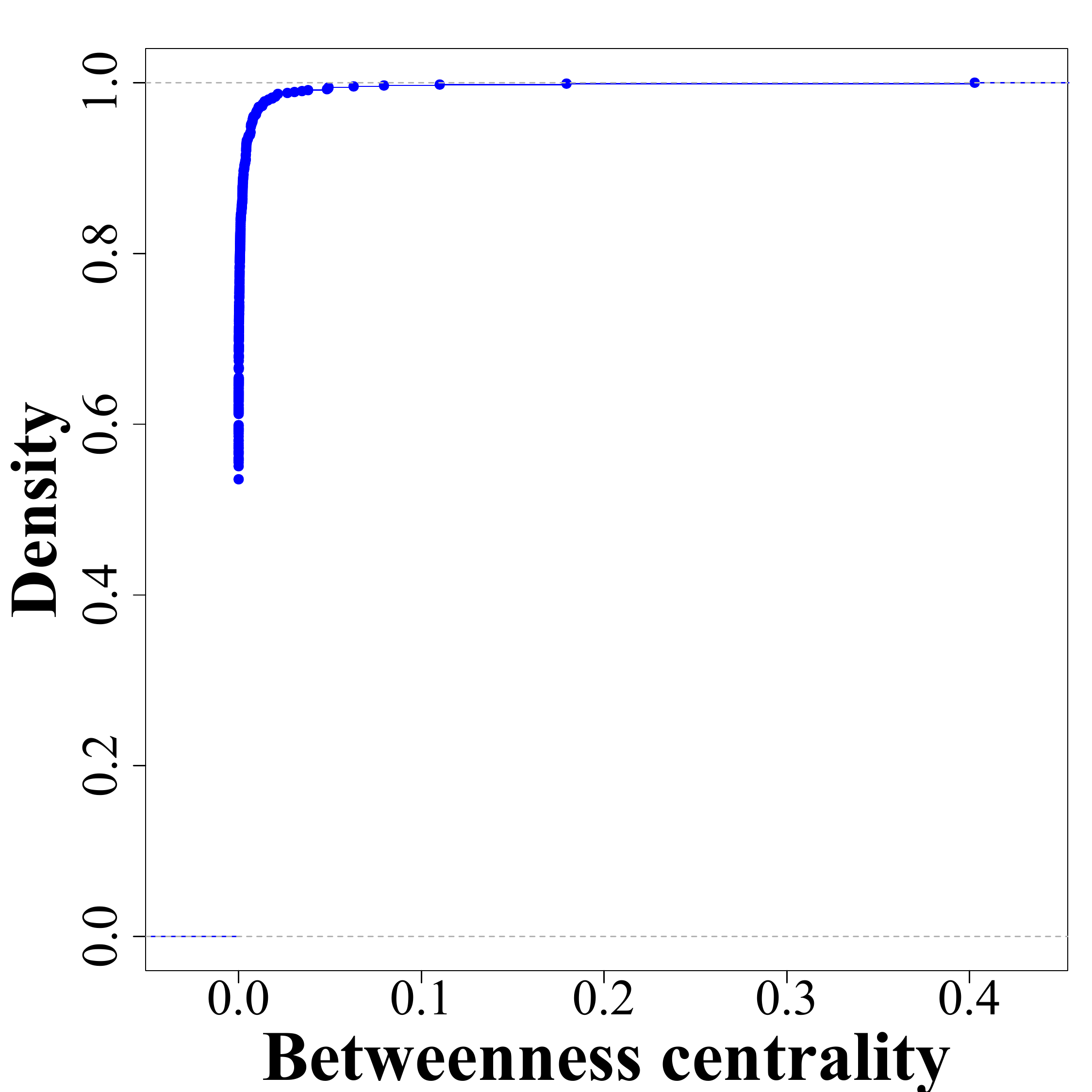}%
	\label{fig:btwn}}
\hfill
	\subfloat[Eigenvector]{\includegraphics[width=0.15\textwidth]{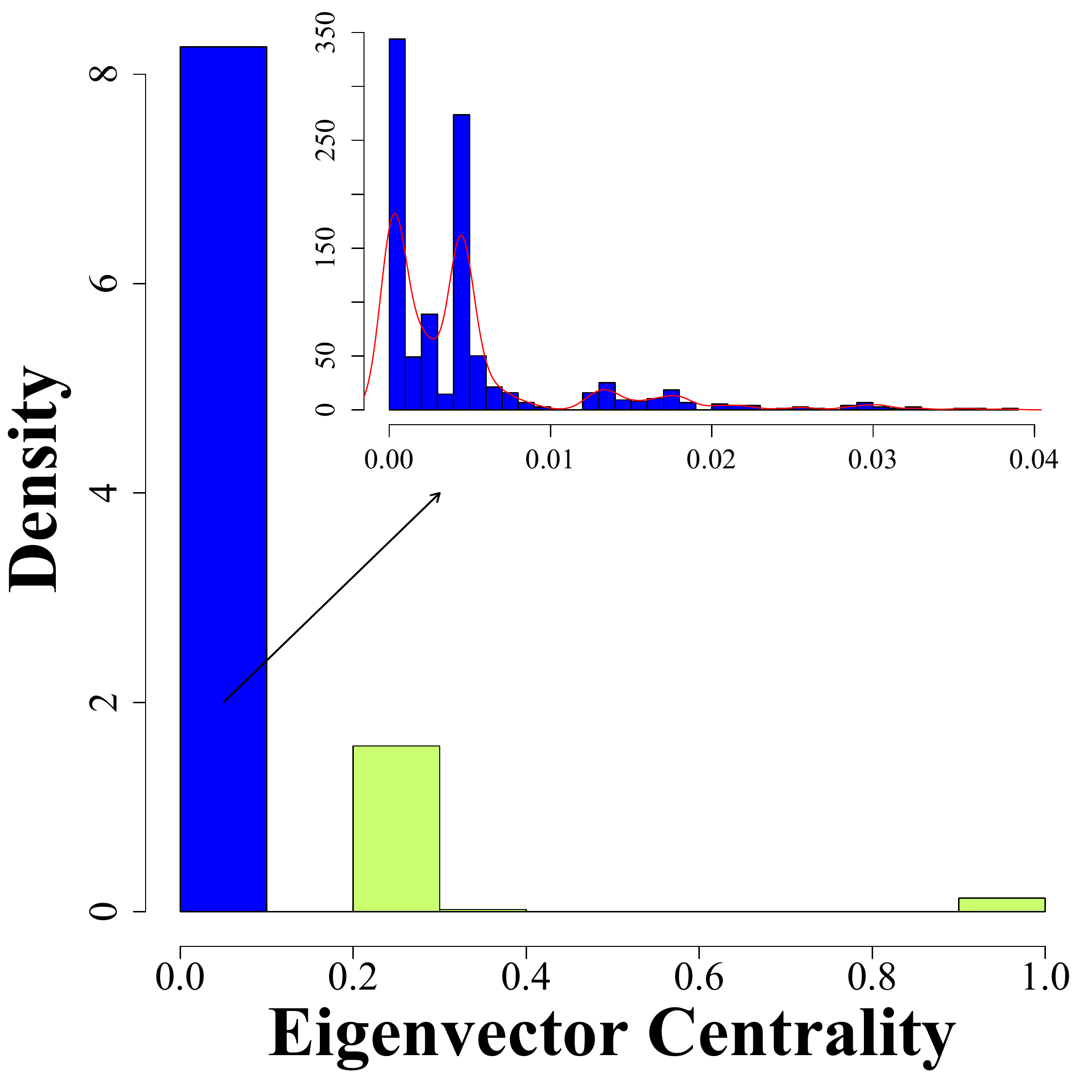}%
	\label{fig:eigen}}
\hfill
	\subfloat[Closeness]{
	\includegraphics[width=0.15\textwidth]{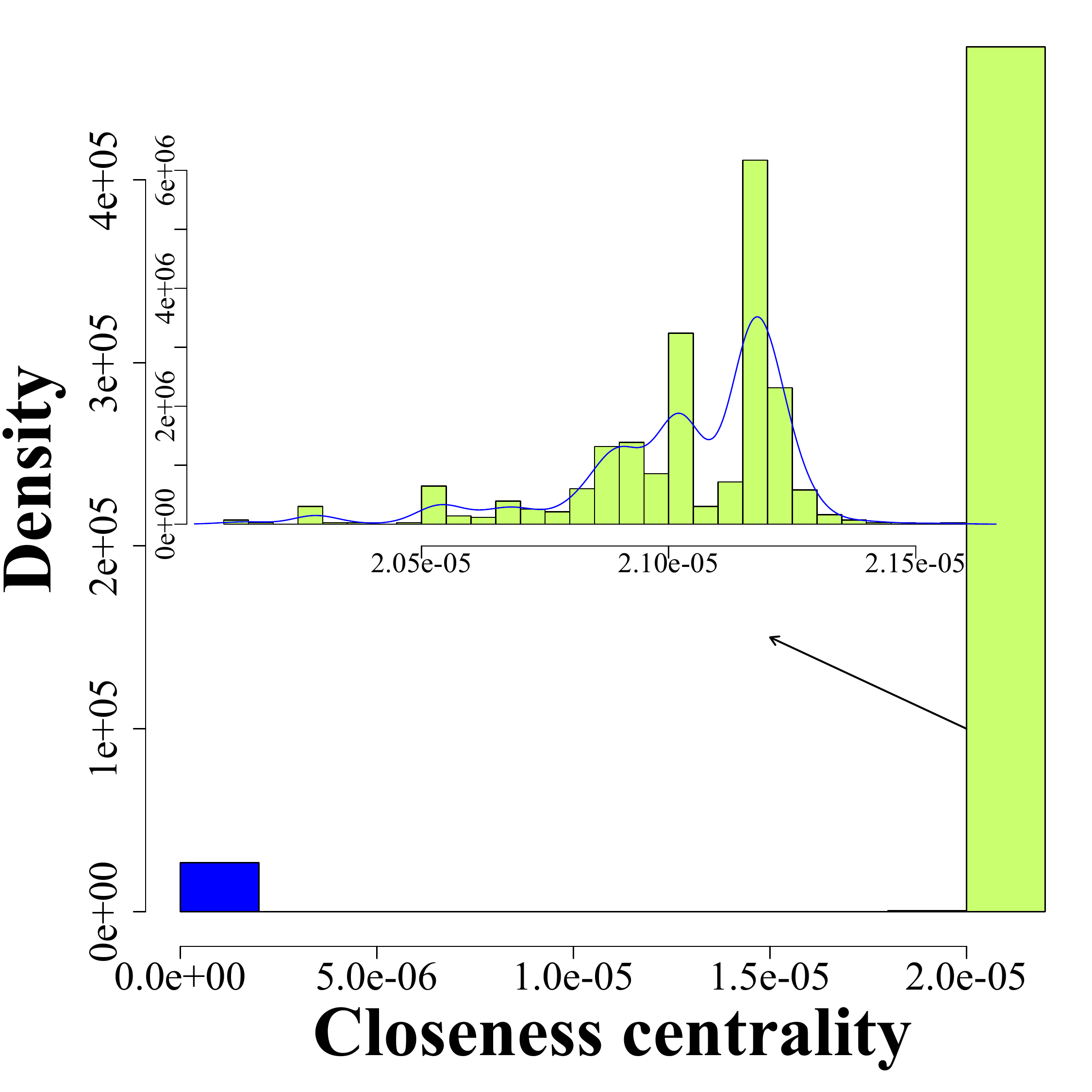}%
	\label{fig:close}}
\caption{Centrality distributions}
  \vspace{-12px}
\end{figure}

\subsection{Importance analysis}
\label{sec:import}

We further examine the ``importance'' of particular domains as defined by various measures of network centrality. The centrality analysis only considers domains having in- or out-degree $\geq 1$.
We show the CDF of betweenness centralities $b_c$ of domains in Figure~\ref{fig:btwn}. 
It shows how virtually every Tor domain has a betweenness centrality near or lower than
$0.05$; in fact there are only $6$ domains with betweenness centrality greater than $0.05$.
These domains are the directory $Hidden Wiki$ ($b_c = 0.4$) 
(matching previous reports suggesting this domain is the principle Tor directory~\cite{cyberSecurity}), 
the Email domain $TorBox$ ($b_c = 0.17$) and $VFEmail$ ($b_c = 0.06$), 
a Dream market domain ($b_c = 0.07$), and
two directories in the $Tor Wiki$ domain ($b_c = 0.19$ and $0.11$ respectively).
An attack, removal, or failure of such directories may thus
directly impact the number of Tor domains reachable by a casual browser exploring 
this dark web. These domains may further be crucial entry points for probes or crawlers
seeking to map the structure of Tor.

Figure~\ref{fig:eigen} shows the distribution of eigenvector centralities across Tor domains. 
We use a histogram rather than a CDF plot to better illustrate the variability  between centrality values. 
Eigenvector centrality~\cite{centrality} describes the structural importance 
of a node as a function of the importance of its neighboring nodes. The 
eigenvector centrailty of vertex $v_i$ is given by the $i^{th}$ component of the 
eigenvector of $A$ whose corresponding eigenvalue is largest. 
We find a heavily skewed distribution of eigenvector centralities where a majority
are close to zero. Further investigation revealed that all domains with
eigenvector centrality $\geq 0.2$ are part of the Dream market. Although high eigenvector centrality does not
correlate with its relative popularity or frequency of visits from users, the naturally developed organization of Tor's domain
structure places the Dream market as the most meaningful Tor domain by a wide margin (note that the highest eigenvector
centrality of a non-Dream market domain is only $0.04$). This establishes the Dream market
as the most structurally important, ``core'' service Tor provides.  
The inlet of Figure~\ref{fig:eigen} gives a sense of the distribution for the remaining domains. 
Here, the distribution exhibits a number of modes corresponding to connected components of the network
that is disconnected from the Dream market. The especially low eigenvector centralities of these 
domains are further indicative of the significance of the Dream market's structural importance
in Tor. 

Figure~\ref{fig:close} shows the distribution of closeness centralities. Like eigenvector centrality, we find a division of domains by those that exhibit 
extremely low or high scores, but here the majority of domains have very high closeness centrality. It is interesting to find that most domains exhibit a high
closeness centrality despite intra-connectivity analysis from Section~\ref{sec:intraConn} suggesting that the intra-connectivity of Tor 
domains is sparse and has a small largest W.C.C. This outcome is likely the product of the directories $Hidden Wiki$ and $Tor Wiki$ having high
betweenness centrality that enables many pairs of domains to be few hops away from each other via these directories. This underscores the central
importance of directory domains to connect Tor pages across domains, and the fact that Tor domains tend to remain undiscoverable without
directories. 

\subsection{Scale-free structure}
\label{sec:scale}
We also investigate signs that the hyperlink structure of domains, and structure within domains, take on the same scale-free structure 
seen in many other sociotechnological systems~\cite{barabasi} including the 
hyperlink structure of the surface web~\cite{connectedWeb}. 
\begin{figure}
\includegraphics[width=0.35\textwidth]{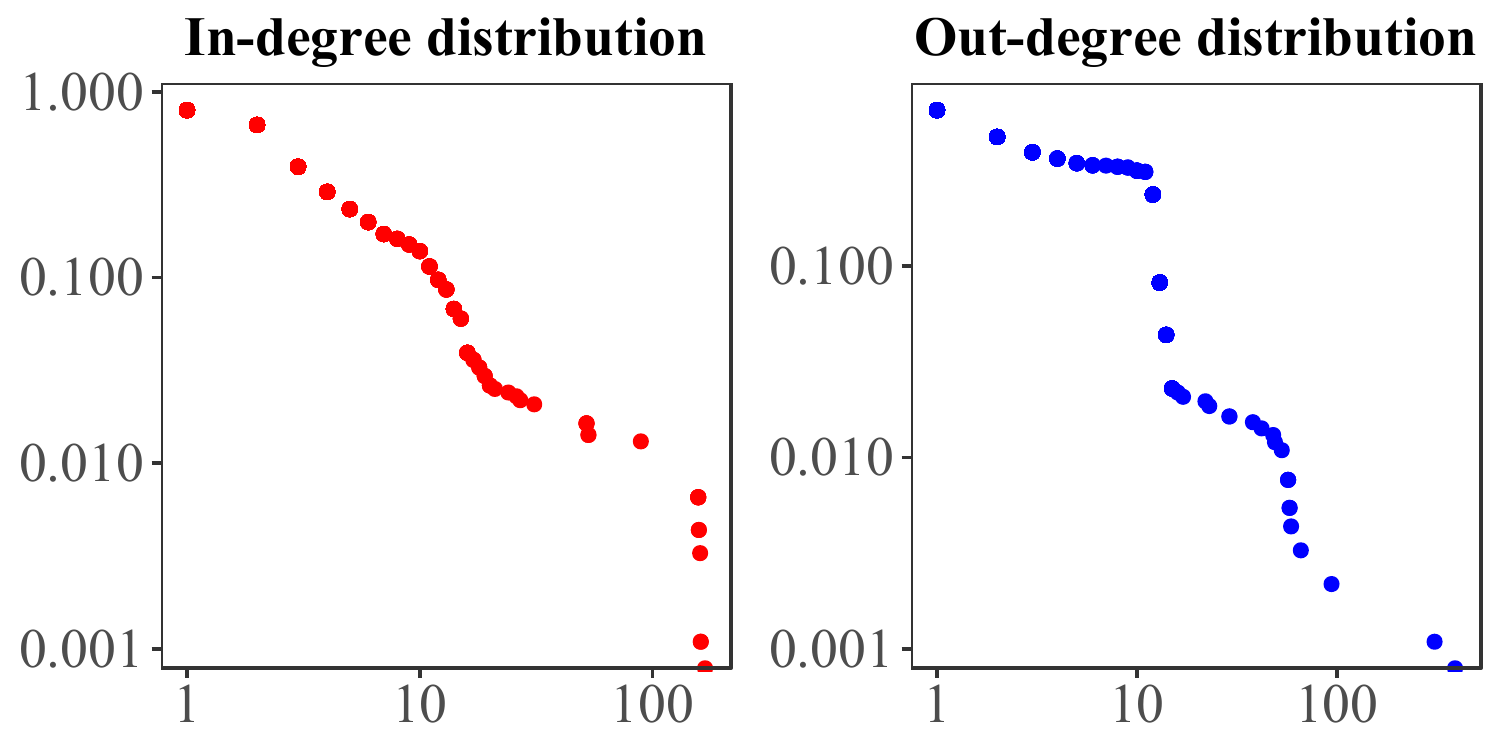}
\caption{CCDF of network degree distributions}
\label{fig:CCDFTotalDegrees}
\vspace{-10px}
\end{figure}
Figure~\ref{fig:CCDFTotalDegrees} show the CCDF of the degree distributions on log-log scale. 
The in-degree distribution does not exhibit a straight line pattern indicative of a power-law, with
a rapid drop in the CCDF occurring in the body of the distribution around an in-degree of 10. 
The out-degree distribution takes on a bimodal pattern, with a set of domains having degree less than
10 and another with degrees between 10 and 100. The distribution's patterns may be explained by the 
variety of inter- and intra-connectivity patterns observed within each of the domains studied in 
Sections~\ref{sec:interConn} and~\ref{sec:intraConn}.
To quantitatively confirm the distribution is not a power-law we 
apply Clauset {\em et al.} hypothesis test presented in~\cite{clauset}.
The test checks the null hypothesis \textit{$H_0$: the network degree distribution is power-tailed}
against the alternative that it is {\bf not} power-tailed, 
and provides an estimate of the power-law exponent $\alpha$ under $H_0$. 
The test leaves little doubt that the in- and out-degree distributions of the network (Figure~\ref{fig:CCDFTotalDegrees}) are not power-tailed
with $p=0.0001$, $p=0.056$, respectively. If we consider the edges to be undirected we still have evidence to reject $H_0$ 
with $p=0.0003$. These measurements confirm the analysis presented in~\cite{onionEyes} that 
the hyperlink structure of Tor domains does not have the same scale-free structure as the surface web. 

Noting the variety of intra-connectivity patterns discussed in Section~\ref{sec:intraConn}, 
we also check if the degree distribution of sites within each domain are power-tailed. These measurements
include intra-domain connections as well as those connections incident to a different domain. We list the 
$p$-value of the test for the in- and out-degree of each domain in Table~\ref{tab:pvalues} and include the 
estimate of $\alpha$ when $H_0$ cannot be rejected. Interestingly, we note that it is only for the
in-degree distributions of Bitcoin, forum, email, shopping, and directory domains, and the out-degree of the news domain, 
where there is insufficient evidence to reject $H_0$. The popularity of about half of all Tor domains (where popularity
is defined by an incoming hyperlink) thus has a power-tailed pattern suggesting that a 
a small number of Bitcoin, forum, email, shopping, and directories are linked to many times more frequently than 
is typical. 
That the news domain is the only one with a power law out-degree distribution may suggest the presence of a 
small number of highly active news sites that offer posts discussing a far wider variety of other Tor domains compared to other news domains. 
The majority of news domains may thus focus on a specific topic, or are otherwise used to discuss events
outside of the Tor network. 
\begin{table}
\vspace{-5px}
\caption{Power-tail distribution hypothesis test results}
\centering
\begin{tabular}{lll}
\toprule
Community&In-degree distribution&Out-degree distribution\\
\midrule
Bitcoin & $p=.7623 , \alpha=2.69$ & $p=.0114$\\[-1pt]
Forum& $p=.8996 , \alpha=3.01$ & $p=.0001$\\[-1pt]
Email& $p=.3377 , \alpha=2.08$ & $p=.0086$\\[-1pt]
News& $p=.0344$ & $p=.5681 , \alpha=2.73$ \\[-1pt]
Directory& $p=.3407 , \alpha=2.65$ & $p=.0002$\\[-1pt]
Shopping& $p=.2021 , \alpha=2.85$ & $p=.0001$\\[-1pt]
Gambling& $p=.0002$ & $p=.0002$ \\[-1pt]
Multimedia& $p=.0003$ & $p=.0002$\\[-1pt]
Dream Market & $p=.0005$ & $p=.0007$\\[-1pt]
\bottomrule
\end{tabular}
\label{tab:pvalues}
\vspace{-17px}
\end{table}

\section{Conclusions and Future Work}
\label{sec:co}
This paper presented a broad overview of the content of English language Tor domains
captured in a large crawl of the Tor network. The paper makes revelations about not the physical or logical (hyperlink) structure of Tor, but of 
the particular domains of information or services hosted on the service and the structure between and within such domains. 
Such birds-eye insights, and especially those related to the inter-connectivity of domains, cannot be acquired by 
synthesizing the existing low-level content analysis work that focuses on a single type of Tor domain. 
 Content analysis carried out by LDA and GbTL, using measures of topic coherence and 
label suitability, identified just nine principal types of domains. Manual analysis of each domain was done to describe the meaning
of each topic, and any standout `sub-types' seen within them.
Over half of all domains constitute site directories or marketplaces
to purchase and sell goods or services, with money tansfer, drugs, and pornography servicing as the most popular types of marketplaces. 
Our measurements identified the Dream market as perhaps the `core' service of Tor, as Dream market domains exhibit especially high
closeness and eignevector centralities. The inter-connectivity of the Tor domain network is surprisingly sparse with a small maximum W.C.C.
but interesting domain inter-connection patterns discussed in Section~\ref{sec:interConn}.
 Patterns in the intra-connectivity structure are further indicative of levels of cooperation 
(where some pages hyperlink to pages in the same domain) and competition (where pages in a domain are more likely to isolate themselves
from pages in the same domain) that may be measured by robustness coefficients $R_\kappa$ for varying centrality scores $\kappa$. 
We further note evidence for rejecting the hypothesis that the global domain structure is scale-free, yet there is insufficient evidence 
in the in-degree distributions of some domain intra-networks and the out-degree distribution of the news domain intra-network to 
conclude that these subnetworks are not power law. This is indicative of different underlying processes that form
connections in different intra-domain networks. 

Future work can expand this study further by replicating our analysis on Tor pages of different languages 
(chinese, russian, persian), and then by studying topic inter-connectivity between language domains, can 
yield a global perspective into the Tor content ecosystem. Such analysis can further be replicated on other dark web services 
to better understand why these lesser known services are used. Another direction of future work is to study the evolution of the topics and communities over time using a richer topic extraction analysis based on 
algorithms discussed in~\cite{hofmann2017,callon1983}. Finally, the study can be combined with a modern crawl of similar domains
on the surface web to be able to directly compare and contrast surface and dark web hyper-link structure. Such a comparative 
analysis could shed light around the differences in use and information between the surface and dark web. 

\begin{acks}
This paper is based on work supported by the National Science Foundation (NSF) under Grant No. 1464104. Any opinions, findings, and conclusions or recommendations expressed are those of the author(s) and do not necessarily reflect the views of the NSF.
\end{acks}

\bibliographystyle{ACM-Reference-Format}
\bibliography{sample-bibliography}

\end{document}